\documentstyle[aas2pp4]{article}

\begin{document}
\title{EUVE Observations of Her X-1 at the End of the Short High State}

\author{D. A. Leahy}
\affil{Dept. of Physics, University of Calgary, University of Calgary,
Calgary, Alberta, Canada T2N 1N4}
\and
\author{H. Marshall} 
\affil{M.I.T., Cambridge, MA 02139}
\authoremail{leahy@iras.ucalgary.ca}
 
\begin{abstract}
Observations of Her X-1 by the Extreme Ultraviolet Explorer (EUVE)
at the end of the x-ray Short High state are reported here. 
Her X-1 is found to exhibit a strong orbital modulation of the 
EUV flux, with a large dip superposed on a broad peak around 
orbital phase 0.5 when the neutron star is closest the observer. 
Alternate mechanisms for producing the observed EUV lightcurve are modeled.
We conclude that: i) the x-ray heated surface of the companion 
is too cool to produce enough
emission; ii) the accretion disk can produce enough emission but does not
explain the orbital modulation; iii) reflection of x-rays off of the companion
can produce the shape and intensity of the observed lightcurve.
The only viable cause for the large dip at orbital phase 0.5
is shadowing of the companion by the accretion disk.
\end{abstract}

\section{Introduction}

      Hercules X-1 is one of the brighter, and most studied, of the
x-ray binary pulsars. It exhibits a wealth of phenomena, including 
pulsations at 1.23 seconds, eclipses at the orbital period of  1.7 day, and
a 35 day cycle in the x-ray intensity. Her X-1 is reviewed by \cite{sco93}.
Recent discussions of the properties of the 35 day cycle are given by
\cite{sco99} and \cite{sha98}. The x-ray pulse profile evolution is discussed in
\cite{dee98}. Recent x-ray spectra of Her X-1 are given by \cite{oos98} and
\cite{dal98} (from BeppoSAX) and \cite{cho97} (from ASCA). An updated set of
binary parameters is given by \cite{lea98}. Analysis of ultraviolet spectra
of Her X-1 are presented by \cite{bor97}, \cite{vrt96}. Optical signatures of
reprocessing on the companion and accretion disk are discussed by \cite{sti97}.
 
Her X-1 has the further advantage of a low interstellar
hydrogen column density, making EUV observations feasible.
Her X-1 has been observed in the EUV previously (\cite{roc94}, 
\cite{vrt94}). 
Here we report the results of analysis of EUVE observations of Her X-1 
covering two complete orbital cycles at the end of the Short
High state of Her X-1.

\section{Observations}

Hercules X-1 was observed with the Extreme Ultraviolet Explorer (EUVE)
on June 24-28, 1995 (TJD = JD-2440000 = 9893.0 - 9897.1). See 
\cite{mal93} for a 
description of the EUVE instruments. Using the 35 day ephemeris from 
BATSE and XTE/ASM observations (Scott, private communication) for phase
0 of 34.85N + MJD50077.0, the 35 day phase of the EUVE observations is
0.71 - 0.82. Phase 0 is defined as turn on of the Main High state.
The shape of the 35  day cycle and durations of the different states has
only recently been well determined (\cite{sco99}): the Main High and
Short High states cover 35 day phase 0-0.33 and 0.59-0.80.
Thus the EUVE observations occur during decline of the Short High 
state into the low state. 

Her X-1 was detected during these observations as a source in  
the Deep Survey (DS) Spectrometer, with Lexan/B filter.
However it is a faint EUVE source so it  did not have enough signal to give a
usable spectrum. 
The DS Lexan/B image shows that Her X-1 is clearly detected.

The DS lightcurve of Her X-1 is given in Figure 1, which 
shows the net source count rate in the Deep Survey instrument for the 
observation period. The error bars are $\pm1\sigma$.  
The orbital modulation at the 1.7 day orbital 
period of Her X-1 is clearly seen. 
The EUV flux is within $2\sigma$ of zero
during the known times of x-ray 
eclipse, e.g. as measured by GINGA (\cite{lea95}).
Binary phase is defined here so that binary phase 0.0 
is center of eclipse of the neutron star by the companion.

Broad dips are seen at TJD (=JD-2450000) 9894.2 and 9895.9.
In addition there are three narrow dips detected in the EUV lightcurve: at 
JD-2440000 of 9894.54, 9895.53, and 9896.19. The minimum points of the
 narrow dips differ from adjacent points by $2.3\sigma$, $3.6\sigma$, and $2.7\sigma$ respectively.
The orbital phases of
these dips is 0.707, 0.301 and 0.684, using the orbital ephemeris of
\cite{sco93}. The first and third dips are pre-eclipse dips and show 
a separation of 1.62 d, The second dip is likely an anomalous dip. 

The EUVE lightcurve of \cite{vrt94} covered a similar time interval (just over
2 binary orbits of Her X-1). The observations occured during the main high state of the
35 day cycle, but were affected by the onset of an anomalous low state. That EUVE
lightcurve had bright phases ($\sim0.6 c/s$) and faint phases ($\sim0.02c/s$). The
lightcurve presented here is comparable in intensity to the faint phases of that
presented by \cite{vrt94}. 

\section{Modelling the EUV Lightcurve}

Modelling calculations were carried out to compare to the observed
EUVE DS lightcurve. 
The orbital parameters from x-ray timing analysis (\cite{dee91}) were used. 
Additional assumptions were needed for the orbit to calculate light curves: 
orbital inclination of 85$^o$, a $K_{opt}$ value of 100 km/s (\cite{rey97}),
and a mass ratio of 0.592 (\cite{lea98}). The distance of Her X-1 was
assumed to be 6.6 kpc.

\subsection{X-ray Heating of HZ Her}

One contribution to the EUVE lightcurve is emission from
the x-ray heated face of the optical companion HZ Her.
That has been calculated here, giving a EUV lightcurve 
that can reproduce the general shape of the
orbital modulation seen in Figure 1, including
the dip at orbital phase 0.5, which is produced by the accretion 
disk occulting HZ Her.
For Her X-1, the column density is $\sim1-5\times10^{19}cm^{-2}$ 
(\cite{mav93}, \cite{vrt94}, \cite{dal98}). When we include this and the
response of the EUVE DS, the model count rate
is smaller than the observed count rate by a large factor. One can achieve a large
enough count rate by artificially raising the temperature of the heated face
of HZ Her by a factor $\sim2$. However this is far larger than any uncertainties
in the calculation (or in the observed temperature of the heated face of HZ Her), 
so the conclusion is that this x-ray heating model fails to
explain the data. 
 
\subsection{Accretion Disk EUV Emission}

The accretion disk has two sources of EUV emission: the x-ray heated surface
(illuminated by the pulsar); 
and the emission from self-heating by viscous dissipation in the disk.

First we consider a simple model for emission from the heated surface of the disk:
a uniform temperature hot spot. The actual size and
shape of the heated region which is visible to the observer depends in detail on the
geometry of the twisted and tilted disk. 
The hot spot is taken a blackbody of circular shape (normal to the line-of-sight),
with radius R and temperature T.  
By requiring the model
count rate (including interstellar absorption and the EUVE DS spectral response) to 
be 0.03 c/s we obtain a single  constraint relating R and T. 
E.g. for a column density of $5\times10^{19}cm^{-2}$, $T= 10^5 K$ gives 
$R=5.9\times10^9cm$, and
$T= 10^6 K$ gives $R=8.5\times10^6cm$.

Next we consider the emission from a viscous disk. In this case we use a model which
includes both viscous heating and x-ray irradiation  (\cite{sch94}). 
This disk has temperature-radius relation:
 $T=1.5\times10^5(R/10^9cm)^{-0.6} K$.  We include interstellar
absorption and then fold an approximation to the spectrum with the EUVE DS spectral
response to determine a model EUVE DS count rate. For a face-on, unobscured
disk face, and a column density of $5\times10^{19}cm^{-2}$, the model count rate
is several times the observed count rate. The model emission comes from the part of
the disk within $10^{9}cm$ of the center, where the disk temperature is more than
$1.5\times10^5K$. 
If one includes the effect of high inclination of Her X-1, and self-occultation
by other regions of the disk due to the
twist and tilt of the disk (e.g. see the disk model of \cite{sch94}), 
the model count rate will be reduced by a factor of 
$\sim 3$ to $>10$, which is sensitive to details of the geometry. 
Thus disk emission is consistent with the observed magnitude of the DS count rate.
 
The disk as source of the EUV emission (either from x-ray heating or viscous heating)
has the following difficulty.
The disk in Her X-1 is generally accepted to precess with a 35 day period. 
The orbital modulation in  disk EUV emission models is restricted to eclipse of 
(small) heated  part of the disk by the companion, which is restricted to small range
of orbital phases between 0.9 and 0.1.  
There is no easy way for the disk models to produce the observed strong
orbital modulation. 

\subsection{Reflection of X-rays from HZ Her}

Another source of EUV emission is the long wavelength part of the 
spectrum of scattered x-rays from the system. 
The x-ray spectrum of Her X-1 below a few hundred eV
is dominated by the blackbody component (\cite{mcc82}, \cite{vrt94}, 
\cite{mav93}, \cite{dal98}).
From the observers vantage point, by far the largest area for 
reflecting x-rays is the illuminated surface of HZ Her.
This illuminated surface of HZ Her has an outer ionized layer 
which reflects x-rays by electron scattering.

We have calculated the amount of reflected x-rays as a function of orbital phase, expressed as a fraction of the direct flux from the neutron star. 
We take the reflecting layer as a thin 
layer located at the Roche critical surface, and assume that a fraction, 
$\eta$, of the incident x-ray flux is scattered isotropically. 
R is taken as independent of energy, so that the spectrum of the
scattered radiation is the same as that of the incident radiation.
The resulting light curve is a smooth function of orbital phase, peaked
at orbital phase 0.5 and zero around orbital phase 0.
The model light curve for $\eta$=0.5 fits the observed data reasonably well, 
except for the region around orbital phase 0.5.  
The value of $\eta$ is fairly high, so indicates that 
the material scattering the x-rays is mostly ionized. However the
value of $\eta$ is not accurate, due to uncertainties in the normalization 
and spectrum of the soft x-ray flux and in the distance to Her X-1.
We estimate an uncertainty in the value of $\eta$ by a factor of $\sim2-4$.

\subsection{Origin of the Dips}

Next we discuss the large dips near orbital phase 0.5.
The observed dips at TJD9894.2 and TJD9895.8 have intensity 
reductions of 73\% and 56\%, resp. 
First we describe a simple model to fit the observed light curves including the
large dips. 
The light curve for reflected intensity, with  $\eta=0.5$,  
was calculated as above, then  multiplied by a shadowing function. 
The shadowing function was of the form:
$(1-\alpha~exp(-(\phi-\phi_o)^2/2\sigma^2))$, with $\phi$ is orbital phase.
The parameters $\alpha$, $\phi_o$ and $\sigma$ were allowed to have different
values for the two observed orbits, and were varied to
achieve a good fit to the data. 
Since there was no improvement in allowing the two values of $\sigma$ to be
different, they were set to be the same, giving  $\sigma=30^o$.. 
The resulting model light curve is shown in Fig. 2 by the solid line, with the
data points plotted as the circles.
The resulting parameter values are:  
$\alpha=0.85$ and $\phi_o=189^o$ for the first orbit;  
$\alpha=0.8$ and $\phi_o=178^o$ for the second orbit. 

Next we discuss the physical origin of the dips. 
One can achieve a reduction in flux by blocking the line-of-sight to HZ Her 
by the accretion disk. 
Calculation shows that the reflected x-rays 
come fairly uniformly from the whole face of HZ Her, 
so one needs an object nearly the same size as HZ Her to achieve the observed
large flux reductions.
(In contrast, for the x-ray heating model, the EUV emission was highly concentrated near 
the L1 point so the accretion disk could easily block the emission). 
The reduction in reflected flux for the case of a spherical occulting surface 
of radius R (centered on HZ Her) is calculated to give an estimate of the required
size of an occulter. 
The results are that the reduction in flux is a smooth function of  $R/R_{HZ Her}$. 
Sample values of the flux reduction are:  
10\% at $R/R_{HZ Her}=0.26$, 50\% at $R/R_{HZ Her}=0.61$,
80\% at $R/R_{HZ Her}=0.83$.

The largest object available for occultation in the system is the accretion disk. 
The radius of the outer edge of the disk is somewhat less than that of the Roche 
lobe of Her X-1, which is at $2\times10^{11}cm$, 
(calculated using the binary parameters of \cite{lea98}).   
\cite{sch94} gives a better limit on the accretion disk radius, 
based on the observed orbital period change, of $1.7\times10^{11}cm$.
The disk model of \cite{sch94} has an outer edge inclination of $\sim7^o$,
which results in a flux reduction of 13\% for the most favorable disk orientation,
assuming a system inclination of $85^o$. 
Thus occultation is not capable of explaining the dips near orbital phase 0.5.

The alternative explanation is that HZ Her is shadowed from the x-ray
source by the accretion disk. 
The accretion disk then only needs to subtend a 
significant angular extent viewed from the neutron star. 
For the geometry of Her X-1, the angular radius of HZ Her viewed from 
the neutron star is $25^o$.
Significant shadowing of this can occur with a twisted 
tilted accretion disk.
An example of a calculated shadow is given by Figure 5 of \cite{sti97}, which
shows $\sim$25\% of the front side of HZ Her shadowed for the particular disk
parameters they have chosen. 
\cite{sch94} give a sketch of a similar disk (their Fig.13), and a depiction of
the shadow the disk casts on the sky as viewed from the neutron star (their Fig.
12).

The amount of shadowing can be calculated from existing disk models. 
The result depends on the disk tilt and on the disk twist.
The maximum tilt is $\sim10^o$ in the model of \cite{sch94}, but
higher in other models (e.g. \cite{sco93} has a maximum tilt of $30^o$). 
Models with a maximum disk tilt of $10^o$ cannot give a flux reduction,
even at most favorable orientation, larger than $\sim30\%$. The larger flux
reduction is associated with models with larger twist. 
A flux reduction of 100\% is possible for maximum tilt greater than $25^o$,
for which case the vertical angular extent of the accretion disk is larger
than that of HZ Her.
The general conclusion is that shadowing can account for the large dips 
near orbital phase 0.5.
However, the observed strength of the dips, 
as large as $\sim$60-70\%, implies that
the maximum disk tilt should be larger than $\sim20^o$. 
 
Further evidence that the dips are due to accretion disk shadowing comes from
the timing of the dips.
The disk in Her X-1 precesses counter to the orbit over a 
35 day period, so the disk shadow moves to earlier orbital phase by $18^o$ 
(or 0.05 in orbital phase) during a single 1.7 day orbit. 
Equivalently, the shadow has a 1.62 day period.
Compare this to the observed dips. 
The separation between the two minimum intensity points 
(at  JD-2440000 of 9894.22 and 9895.86 in Figure 1) 
is 1.64 day, with an uncertainty of $\sim 0.2$ day.
(A different way of measuring the same offset is by use of 
the Gaussian shadow model. It yielded an orbital phase difference between the
shadows for the two orbits of $11^o$.)
Thus the observed dips in the light curve are consistent with an origin in the
precessing shadow of the accretion disk, but not consistent
with a constant period of 1.7 day.

Next we discuss what shadowing is expected from standard disk models as a
function of 35 day phase, and in
particular, at the 35 day phase of the observations here.
The timing (i.e. orbital phase) 
of the accretion disk shadow can be predicted from the 35 day phase since
they both depend on the orientation of the accretion disk with respect to the
observer. 
The Main High state peaks near 35 day phase 0.12 (\cite{sco99}) so the observer
experiences minimum accretion disk blockage at this time. 
So the shadow on HZ Her should be minimum 
at orbital phase 0 (when HZ Her experiences the shadow closest to the 
direction of the observer) closest to  35 day phase 0.12. 
However we observe the shadow near orbital phase 0.5. To get the
shadow at orbital phase 0.5 we just need to rotate the disk by $180^o$, 
i.e. change 35 day phase by 0.5.
Thus the shadow on HZ Her, at orbital phase 0.5, is minimum  at 35 day phase 0.62. 

The 35-day phase of maximum of disk shadow to the observer depends on the details
of the disk model.  
Tilted-twisted disk models have rings at each radius which cross the
binary plane twice over $360^o$ in azimuth, with an assumed symmetry that
has crossings (and maximum excursions from the binary plane) separated by $180^o$.
This results in maximum shadow at 35 day phases $\phi$ and  $\phi+0.5$.  
In the model of \cite{sch94}, the maximum shadow follows minimum
shadow by 0.23 in 35 day phase (compared to 0.22 for the model of \cite{sco93}).  
Thus the shadow should peak at orbital phase 0.5 in this model 
at 35-day phase 0.85.
 
From the orbital phases of the two observed dips 
we estimate the time that the shadow maximum occurs at orbital phase 0.5. 
This yields a time of JD-2440000 = 9895 or 35 day phase 0.77,
which is different from the above prediction of 0.85. 
The difference could occur for two reasons:
1. The disk twist and tilt are significantly different than in 
existing disk models,  
so that maximum follows minimum shadow by only 0.15 in 35-day phase
($55^o$ in azimuth). 
2. The minimum obscuration to the observer at 35 day phase 0.12 is offset from the
minimum shadowing of HZ Her at orbital phase 0. 
The latter is hard to achieve, because the observer's 
inclination to the binary plane is only $\sim5^o$ (\cite{lea98}). 
Thus we have evidence here for altered disk parameters
which result in a smaller separation of maximum after minimum shadow.

Twisted-tilted disk models with the symmetry described above
result in a shadow on HZ Her which
repeats twice over the 35 day cycle (except for an inversion about the binary
plane). Thus a prediction is that at orbital phase 0.5 and 35-day phase 0.12 the
shadowing should be a minimum also. 
However the EUVE observations of \cite{vrt94} at this phase
show much brighter EUV emission which is pulsed and
comes directly from the region of the 
pulsar. 
So the much fainter reflected emission off of HZ Her cannot be observed at this phase. 

The three narrow dips   at 
JD-2440000 of 9894.54, 9895.53, and 9896.19  are
interesting. More observations will be needed to verify their
existence. However, if they are verified, they imply a moving structure (with respect to
the disk)
near the neutron star is causing the shadow.  The reason is that a significant (large angular extent)
region of HZ Her must be shadowed, but the shadow must move rapidly compared to orbital period for the
dip to be of short duration compared to the orbital period. Since two of the narrow dips
recur at the same period and orbital phase as pre-eclipse dips, this structure may be 
the same structure that causes the pre-eclipse dips.

\section{Conclusions}

Her X-1 has been detected by the EUVE DS at the end of the Short High state 
of the 35 day x-ray cycle (35 day phase 0.71 to 0.82). 
We have carried out modelling in order to understand the EUVE DS
light curve. The first model considered
was a calculation of the x-ray heating effect from Her X-1 on HZ Her,
including occultation of the heated surface by the accretion disk.
This model can produce the shape of the observed light curve, but 
is of too low intensity mainly due to the effect of interstellar absorption 
at EUV wavelengths.

The next model considered was emission by the accretion disk,
either by a hot spot or by emission from a disk model including viscous
heating and x-ray irradiation.  
The level of observed emission can be produced, and
 is sensitive to the details of the disk geometry.
 However the main difficulty of explaining the observed
EUV emission with a disk is that the disk emission should have a 35 day modulation
and not a modulation at the orbital period, as observed.

The final model considered was reflection of the soft x-ray emission from
the pulsar off of HZ Her. This can give the correct level of emission, and
also produce the observed orbital modulation of the light curve.
The dip near orbital phase 0.5 is produced by shadowing of HZ Her by 
the accretion disk. Further support for this model comes from the difference
in the times of maximum shadow: it is $1.64\pm0.2$ day, consistent with the
expected beat period of 1.62 day between the 1.7 day orbital period and the 35 day
disk precession period, but not consistent with the 1.7 day orbital period.  

The current observations indicate that we do not directly see the soft x-ray pulse at
any time during our observations. In contrast, the EUVE observations of \cite{vrt94}
had bright phases during which pulsations were detected and during which the spectrum
matched that of the soft x-ray pulse. The soft x-ray pulse may come from reprocessing
at the inner edge of the accretion disk. In that case this inner edge must be hidden
from view during our observation (end of short high state), yet visible during much
of the main high state. This is consistent with ideas for explaining the evolution
of the hard x-ray pulse (\cite{dee98}). The faint part of the EUVE lightcurve during
main high shows modulation which may have the same origin as proposed here for 
end of short high. However the bright phases (see Fig.1 of \cite{vrt94}) prevent
one from having a good view of this orbital modulation.

Further observations to determine the EUV light curve at other 35 day phases
will be highly valuable in testing the tilted disk model for Her X-1, and will provide
an opportunity to do detailed modelling of the accretion disk geometry based
on the observed shadowing.
 
\acknowledgments

This work was supported in part by the Natural Sciences and Engineering Research 
Council of Canada.

\figcaption[fig1.ps]
{Observed EUVE DS lightcurve of Her X-1.}
\label{fig.1}

\figcaption[fig2.ps]
{X-ray reflection lightcurve (solid line) for $\eta$=0.5 with the Gaussian shadow
model (see text). 
The  EUVE DS data points are shown by the small circles.}
\label{fig.2}

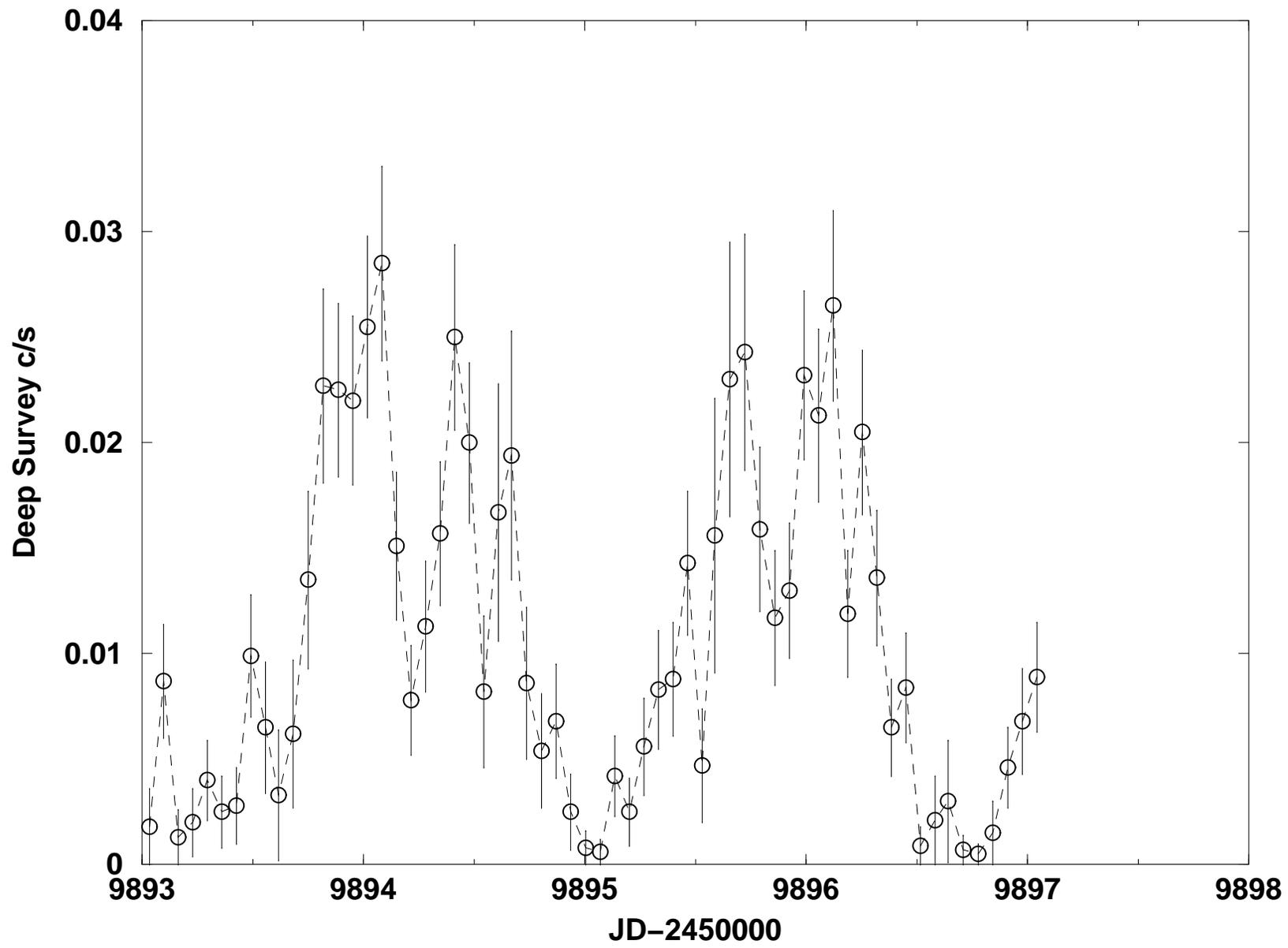
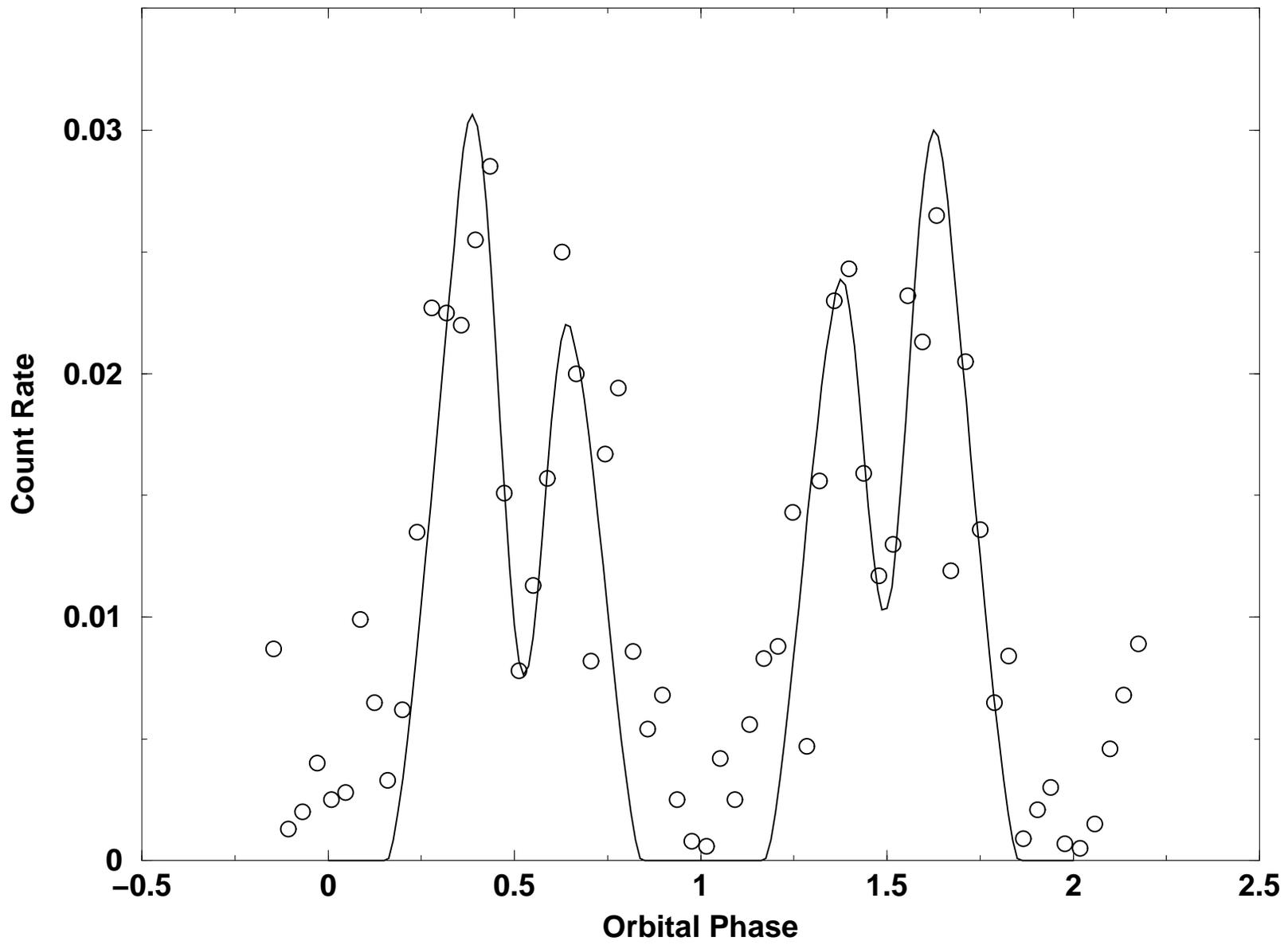
\end{document}